\documentclass[referee]{aa} 
\usepackage{graphics,epsfig} 

\newcommand\lesssim{\stackrel{<}{\sim}} 
 
\begin{document} 
 
\thesaurus{12(12.03.4;  
              12.12.1)} 
     
\title{Limits of Crystallographic Methods for Detecting Space Topology} 
 
\author{Roland Lehoucq\inst{1}  
        \and Jean--Philippe Uzan\inst{2,3} 
        \and Jean--Pierre Luminet\inst{4}} 
\offprints{Roland Lehoucq} 
 
\institute{CE-Saclay, DSM/DAPNIA/Service d'Astrophysique,  
           F-91191 Gif sur Yvette cedex, France\\ 
           email: roller@discovery.saclay.cea.fr 
        \and 
           Laboratoire de Physique Th\'eorique, CNRS--UMR~8627,  
           B\^at. 210, Universit\'e Paris XI, F-91405 Orsay  
           cedex, France\\ 
           email: uzan@th.u-psud.fr 
        \and 
          D\'epartement de Physique Th\'eorique, Universit\'e de Gen\`eve, 
          24 quai E. Ansermet, CH-1211 Geneva (Switzerland). 
        \and 
          D\'epartement d'Astrophysique Relativiste et de Cosmologie, 
          Observatoire de Paris, CNRS--UMR 8629, F-92195 Meudon, France\\ 
          email: Jean-Pierre.Luminet@obspm.fr} 
 
\titlerunning{Limits of Crystallographic Methods} 
\authorrunning{R. Lehoucq, J-P. Uzan \& J-P. Luminet} 
 
 
\maketitle 
 
\begin{abstract} 
We investigate to what extent the cosmic crystallographic methods aimed 
to detect the topology of the universe using catalogues of cosmic 
objects would be damaged by various observational uncertainties. We find 
that the topological signature is robust in the case of Euclidean 
spaces, but is very fragile in the case of compact hyperbolic spaces. 
Comparing our results to the presently accepted range of values for the 
curvature parameters, the best hopes for detecting space topology rest 
on elliptical space models. 
 
 
\noindent{\bf Preprint:} LPT--ORSAY 00/51; UGVA-DPT 00/05--1082
\end{abstract} 
 
%
\section{Introduction}  
  
The search for the topology of the spatial sections of the universe has 
made tremendous progress in the past years (see Lachi\`eze--Rey and 
Luminet (1995) for an introduction to the subject and early references, 
 Luminet and Roukema (1999) and Uzan {\em et al.} (1999b) for a 
review of the late developments). Methods using two--dimensional data 
sets, such as the cosmic microwave background maps planned to be 
obtained by the MAP and Planck Surveyor satellite missions, and 
three--dimensional (3D) data sets, such as galaxy, cluster and quasar 
surveys with redshifts, have been developped. 
  
Following our previous works (Lehoucq {\em et al.} 1996, Lehoucq {\em 
et al.} 1999 and Uzan {\em et al.} 1999a), we focus on the 
topological information that can be extracted from a 3D catalogue of 
cosmic objects. The key point of all the topology--detecting methods is 
based on the ``topological lens effect", i.e. on the fact that if the 
spatial section of the universe has at least one characteristic size 
smaller than the spatial scale of the catalogue, then one should observe 
different images of the same object.  The original {\it crystallographic 
method} (Lehoucq {\em et al.} 1996) used a pair separation histogram 
({\sl PSH}) depicting the number of pairs of catalogue's objects having 
the same three--dimensional separations in the universal covering space, 
with the idea that spikes should stand out dramatically at 
characteristic lengths related to the size of the fundamental domain and 
to the holonomies of space. However, following a remark by Weeks (1998), 
we proved (Uzan {\em et al.} 1999a and Lehoucq {\em et al.} 1999) 
that sharp spikes can emerge in the {\sl PSH} only if the holonomies of 
space are Clifford translations -- a result independently derived by 
Gomero {\em et al.} (1998).  As a consequence, the {\sl PSH} method does 
not apply to the detection of topology in universes with hyperbolic spatial sections. 
  
Since then, various generalisations of the crystallogra\-phic method 
were proposed. Fagundes and Gaussman (1998) suggested to map the 
differences between the {\sl PSH } of a simulated catalogue in a 
compact hyperbolic universe and the {\sl PSH} in the corresponding 
simply--connected universe having the same distribution of objects and 
cosmological parameters. They noticed sharp oscillations on the scale 
of the bin width, modulated by a broad oscillation on the scale of the 
curvature radius. However in Uzan {\em et al.} (1999b), we calculated 
on one hand the differential {\sl PSH } between a simulated 
distribution in a compact hyperbolic Weeks space and a similar 
distribution in the simply--connected hyperbolic space $\bf{H}^3$, on 
the other hand the differential {\sl PSH } between two different 
distributions (with the same total number of objects) in the 
simply--connected hyperbolic space $\bf{H}^3$. Both curves (fig.~10 
of Uzan {\em et al.} 1999b) exhibit the same pattern of sharp and broad 
oscillations, which shows that the topological significance of such a 
pattern is highly doubtful. 
  
Fagundes and Gaussman (1999) next proposed a modified crystallographic  
method in which the topological images in simulated catalogues are  
pulled back to the fundamental domain before the set of 3D distances  
is calculated.  The distribution of pair distances is expected to be  
peaked around zero.  The main drawback of this method is that the  
nature of the signal strongly depends on the topological type, on the  
orientation of the fundamental domain and on the position of the  
observer inside the latter.  Thus the pullback method can be useful  
only if the exact topology is already known.  On the other hand,  
Gomero {\em et al.} (1999) introduced a {\sl mean PSH}, aimed to  
reduce the statistical noise that may mask the topological signature.   
They claimed that such a technique should detect the contribution of  
non-translational isometries to the topological signal, whatever the  
curvature of space, but the applicability of their method to real  
data has not been demonstrated.

In order to improve the signal--to--noise ratio inherent to 
{\sl PSH's }, instead of reducing the statistical noise we can enhance 
the topological signal by collecting all distance correlations into a 
single index. Such was our purpose in Uzan {\em et al.} 1999a, where we 
reformulated cosmic crystallography as a collecting-correlated-pairs 
method ({\sl CCP}). The {\sl CCP} technique rests on the basic fact that 
in a multiply connected universe, equal distances appear more often than 
just by chance, whatever the curvature of space and the nature of 
holonomies.  We also showed that the extraction of a topological signal 
drastically depends on the rather accurate knowledge of the cosmological 
parameters, namely the density parameter, $\Omega_0$, and the 
cosmological constant parameter, $\Omega_{\Lambda0}$. This is due to the 
fact that all cosmic crystallography methods require the determination 
of 3D separations between two any images: observations use 
redshifts for determining the coordinate distance in the universal 
covering space (in addition to the angular positions on the celestial 
sphere), and the redshift--distance relation involves the cosmological  
parameters.  Conversely, the detection of a topological signal would help 
to determine accurately the curvature parameters (see also 
Roukema and Luminet 1999). 
  
However it is to be recognized that, in the framework of cosmic 
crystallography, the application of numerical simulations to real data 
rest on two idealized assumptions, namely: 
\begin{enumerate}  
\item all objects are strictly comoving  
\item the catalogues of observed objects (quasars, galaxy clusters) are complete.  
\end{enumerate}  
  
Although quoted in Lachi\`eze--Rey and Luminet (1995), the 
quantitative effects of such simplifications have not been fully 
discussed in the literature. In Lehoucq {\em et al.} (1996), the 
authors qualitatively argued that the distortion due to peculiar 
velocities was negligible, and they performed numerical simulations to 
study the influence of the angular resolution of the surveys.  In 
Roukema (1996), the influence of the astrophysical uncertainties 
(mainly of spectroscopic measurements and of peculiar velocities) on a 
method trying to find quintuplets of quasars with the same geometry 
(and thus which may be topological images) was evaluated. It was shown 
that, in such a case, the most serious uncertainty comes from the 
radial peculiar velocities. In Uzan {\em et al.} (1999a), we discussed 
the effect of the peculiar velocities and of the errors in the spatial 
position arising from the imprecisions on the values of the 
cosmological parameters. To finish, the errors on the determination of 
the position and of the peculiar velocities were 
discussed in Roukema (1996), and the way that the 
constraints on $\Omega_0$ and $\Omega_{\Lambda0}$ depend on the 
redshifts of multiple topological images and on their radial and 
tangential separations was calculated in Roukema and Luminet (1999). 
  
The goal of the present article is to have a critical attitude on the 
methods we have developped so far, by listing all the sources of 
observational uncertainties and by evaluating their effects on the 
theoretical efficiencies of the {\sl PSH } method (Lehoucq {\em et 
al.} 1996) and of the {\sl CCP} method (Uzan {\em et al.} 1999a) 
respectively.  In \S~\ref{2}, we discuss the nature of uncertainties 
and in \S~\ref{3} the numerical methods used to estimate their effects 
on the topological signal. We then quantify the magnitude of each 
effect in the Euclidean and hyperbolic cases (\S~\ref{4}) (we postpone 
the case of elliptic spaces to a further study). In conclusion 
(\S~\ref{5}) we compare our results to the performances of current and 
future observational programs aimed to detect the topology of the 
universe. \\ 
  
{\bf Notations and descriptions} 
 
We keep the notations of our previous articles (Lehoucq {\em et al.} 
1999, Uzan {\em et al.} 1999a). The local geometry of the universe is 
described by a Friedmann--Lema\^{\i}tre metric 
\begin{equation}  
{\rm d}s^2=-{\rm d}t^2+a^2(t)\left[{\rm d}\chi^2+f^2(\chi) 
\left({\rm d}\vartheta^2+\sin^2\vartheta {\rm d}\varphi^2\right)\right],  
\end{equation}  
where $a$ is the scale factor, $t$ the cosmic time, $\chi$ the 
comoving radial distance, and $f(\chi)=(\sin\chi,\chi,{\rm sinh}\chi)$ 
according to the sign of the space curvature $k=(+1,0,-1)$. 
The time evolution of $a$ is obtained by solving the Friedmann 
and the conservation equations 
\begin{eqnarray}  
H^2&=&\kappa\frac{\rho}{3}-\frac{k}{a^2}+\frac{\Lambda}{3},\\  
\dot\rho&=&-3H(\rho+P)  
\end{eqnarray}   
where $\kappa\equiv8\pi G/c^4$, $\rho$ and $P$ are respectively the  
matter density and the pressure, $\Lambda$  
is the cosmological constant and $H\equiv \dot a/a$ is the Hubble  
parameter (with a dot referring to a time derivative). We also use  
the standard parameters  
\begin{equation}  
\Omega\equiv\frac{\kappa\rho}{3H^2}\quad\hbox{and}\quad  
\Omega_\Lambda\equiv\frac{\Lambda}{3H^2}.  
\end{equation}  
This completely specifies the properties and the dynamics of the  
universal covering space. In the following, we assume that we are in the  
matter dominated era, so that $P=0$ and $\rho\propto a^{-3}$.  
  
The topology of the spatial sections is described by the fundamental 
domain, a polyhedron whose faces are pairwise identified by the 
elements $g$ of the holonomy group $\Gamma$ (see Lachi\`eze--Rey and 
Luminet (1995) for the details). 
 
\section{Observational uncertainties}\label{2}  
  
Early methods used to put a lower bound on the size of the universe  
and based on the direct recognition of multiple images of given  
objects -- such as our Galaxy (Sokolov and Schvartsman 1974), the Coma  
cluster (Gott 1980) -- had to face a major drawback due to the fact  
that the same object would be seen at different lookback times.  Thus,  
evolution effects such as photometric or morphologic changes will, in  
most cases, render the identification of objects impossible (see  
however Roukema and Edge, 1997).  Statistical techniques such as the  {\sl PSH} method (Lehoucq {\em et al.} 1999) and the {\sl CCP} method  
(Uzan {\em et al.} 1999a) are free from such biases. 
  
The two main sources of uncertainties for all the statistical methods 
trying to detect the topology of the universe in catalogues of cosmic 
objects are 
\begin{enumerate}  
\item[(A)] the errors in the positions of observed objects,  
which can be separated into:  
  \begin{enumerate}  
	 \item[(A1)] the uncertainty in the determination of the redshifts  
	 due to spectroscopic imprecision; such an effect is purely  
	 experimental and exists even if the objects are strictly comoving  
	 \item[(A2)] the uncertainty in the position due to peculiar  
	 velocities of objects, which induce peculiar redshift corrections  
	 \item[(A3)] the uncertainty in the cosmological parameters, which  
	 induces an error in the determination of the radial distance (via  
	 the redshift -- distance relation)  
	 \item[(A4)] the angular  
	 displacement due to gravitational lensing by large scale  
	 structure. 
\end{enumerate}  
\item[(B)] the incompleteness of the catalogue, which has two  
main origins:  
\begin{enumerate}  
  \item[(B1)] selection effects implying that   
  some objects are missing from the catalogue  
  \item[(B2)] the partial coverage of the celestial sphere, due either   
  to the presence of the galactic plane, or to the fact that surveys are   
  performed within solid angles much less than $4\pi$.  
\end{enumerate}  
\end{enumerate}  
 
{\bf Effects of peculiar velocities} 
  
A peculiar velocity has two effects on the determination of the position of a 
cosmic object: 
\begin{enumerate}  
\item[(i)] An {\it integrated} effect, coming from the fact that  
if a galaxy has a proper velocity, then its true position differs from its   
comoving one.  
\item[(ii)] An {\it  
instantaneous} effect, due to the fact that the radial component of  
the proper velocity will add to the cosmological redshift an extra term, $\Delta z_{Dop}$.  
\end{enumerate}  
  
Assuming that the vector velocity $\vec v$ of a galaxy is constant, 
which is indeed a good enough approximation to estimate the effects of  
peculiar velocities, a galaxy at redshift $z$ has moved 
from its comoving position by a comoving distance 
\begin{equation}\label{dl}  
\delta\vec\ell=\vec v\tau[z]  
\end{equation}  
where $\tau[z]$ is the look--back time, obtained by integrating the  
photon geodesic equation  
\begin{equation} \label{tau_de_z} 
\tau[z]=\frac{1}{H_0}\int_{\frac{1}{1+z}}^1\frac{{\rm d} 
x}{\sqrt{\Omega_{\Lambda0}  
x^2+(1-\Omega_0-\Omega_{\Lambda0})+\Omega_0/x}}.  
\end{equation}  
This reduces to the well known expression  
\begin{equation}  
\tau[z]=\frac{2}{3H_0}\left(1-\frac{1}{(1+z)^{3/2}}\right)  
\end{equation}  
when $\Omega_0=1$ and $\Omega_{\Lambda0}=0$.

\begin{figure}   
\centerline{  
\epsfig{figure=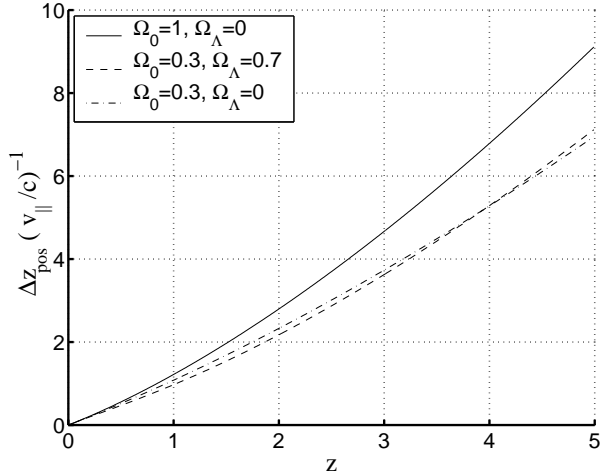,width=8cm}}  
\caption{Variation of $\Delta z_{\rm pos}$ in units  
of $(v_\parallel/c)$ as a function of the redshift, assuming the galaxy   
peculiar velocity is constant.}  
\label{plot_dz}   
\end{figure}

\begin{figure}   
\centerline{  
\epsfig{figure=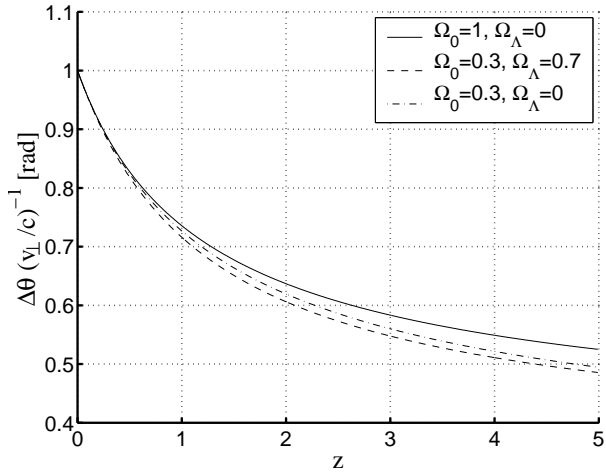,width=8cm}}  
\caption{Variation of $\Delta\theta_{\rm pos}$ as a function of the 
redshift. As long as $z<5$, we have that $\Delta\theta_{\rm 
pos}/1\,{\rm rad}<0.5(v_\perp/c)$. Note also that when $z\rightarrow0$, 
$\Delta\theta_{\rm pos}/1\,{\rm rad}\rightarrow(v_\perp/c)$.} 
\label{plot_dtheta}   
\end{figure}

Now, $\delta\vec\ell$ can be expressed in terms of a peculiar redshift, $\Delta  
z_{\rm pos}$, and of an angular displacement, $\Delta\theta_{\rm  
pos}$,  
which depend on the galaxy velocity as  
\begin{eqnarray}  
\Delta z_{\rm pos}&=&\frac{v_\parallel}{c}\frac{c\tau[z]}{\chi'[z]}\\  
\Delta\theta_{\rm pos}&=&\frac{|\vec v_{\perp}|}{c}\frac{c\tau[z]}{\chi[z]}  
\end{eqnarray}  
 where a prime refers to derivative with respect to $z$,   
$v_\parallel$ is the component of the velocity  
$\vec v$ with respect to the line--of--sight direction $\vec\gamma$, and  
$\vec v_\perp$ is the non--radial peculiar velocity, defined  
as  
\begin{equation}  
v_\parallel\equiv\vec v.\vec\gamma\quad\hbox{and}\quad  
\vec v_\perp\equiv\vec v-v_\parallel\vec\gamma.  
\end{equation}  
$\chi[z]$ is the observer area distance, given by  
\begin{equation}  
\chi[z]=\frac{c}{a_0H_0}\int_{\frac{1}{1+z}}^1  
\frac{{\rm d}x}{\sqrt{\Omega_{\Lambda0}  
x^4+(1-\Omega_0-\Omega_{\Lambda0})x^2+ \Omega_0 x}}.  
\end{equation}  
 In figures  
\ref{plot_dz} and \ref{plot_dtheta}, we respectively depict the  
variations of $\Delta z_{\rm pos}$ and $\Delta\theta_{\rm pos}$ as a  
function of the redshift.  
  
Concerning the instantaneous effect ii), the redshift uncertainty 
$\Delta z_{\rm Dop}$ can be related to the galaxy proper velocity as follows.  
  
If we consider the trajectory of a photon $x^\mu(s)$, $s$ being the 
affine parameter along the null geodesic, the relation between the 
emission (E) wavelength $\lambda_{\rm E}$ and the reception (R) 
wavelength $\lambda_{\rm R}$ can be expressed as 
\begin{equation}  
\frac{\lambda_{\rm R}}{\lambda_{\rm E}}= 
\frac{\left(k^\mu u_\mu^{\rm obs}\right)_{\rm E}}  
{\left(k^\mu u_\mu^{\rm gal}\right)_{\rm R}}  
\end{equation}  
where $k^\mu\equiv dx^\mu/ds$ is the tangent vector to the photon  
geodesic; $u_\mu^{\rm obs}$ and $u_\mu^{\rm gal}$ are respectively the  
4--velocity of the observer and of the galaxy. Neglecting the  
perturbations of the metric and of the matter and focusing on the Doppler  
effect, one can easily show that  
\begin{equation}  
k^\mu u_\mu=-\frac{1}{a(t)}\left[1+\gamma^i\frac{v_i}{c}\right],  
\end{equation}  
where $\gamma^i$ is the direction in which the galaxy is observed and  
$v^i$ is its (Newtonian) velocity. The observed (spectroscopic) redshift,  
$z_{\rm obs}$, and the cosmological redshift, $z_{\rm cosm}$, respectively  
defined by  
\begin{equation}  
1+z_{\rm obs}\equiv\frac{\lambda_R}{\lambda_E},\quad  
1+z_{\rm cosm}\equiv\frac{a(t_E)}{a(t_R)}  
\end{equation}  
are thus related by   
\begin{equation}  
1+z_{\rm obs}=(1+z_{\rm cosm})\left[1+\gamma^i\frac{(v_i^{\rm gal}-v_i^{\rm  
obs})}{c}\right].   
\end{equation}  
Assuming that we can substract the Earth's velocity, an error $\Delta
z\equiv z_{\rm obs}-z_{\rm cosm}$ in the determination of $z_{\rm obs}$
can be interpreted in terms of a radial peculiar velocity given by
\begin{equation}  
\Delta z_{\rm Dop}=\frac{v_\parallel^{\rm gal}}{c}(1+z_{\rm cosm}).  
\end{equation}  
As expected, a galaxy receding from us, i.e. with $\gamma^iv_i^{\rm  
gal}>0$ will have an additional redshift, whereas a galaxy drawing nearer to  
us will induce a Doppler blueshift correction.\\ 
 
The considerations above are also useful for discussing the effects of
angular displacement $\Delta\theta_{\rm gl}$ due to gravitational
lensing by large scale structure (Mellier 1999, van Waerbeke {\em et
al.} 2000).  A typical value $\Delta\theta_{\rm gl} \sim 1$ arcsec
corresponds to a peculiar velocity a few km.s$^{-1}$ as long as the
redshift is less than 5 (as can be seen from
Fig. \ref{plot_dtheta}). Thus effect $(A4)$ is much less important than
effect $(A2)$.

{\bf Effects of catalogue incompleteness} 
 
Objects can be missing from a catalogue due to strong evolution  
effects (this is particularly the case for quasars), and to absorption  
of light by intergalactic gas or dust in some sky directions.  
Another well known selection effect is the Malmquist bias: statistical 
samples of astronomical objects which are limited by apparent magnitude 
have mean absolute magnitudes which are different than those of 
distance--limited samples. An apparent magnitude--limited sample 
contains, if luminosity function has a finite width, some very luminous 
objects which, in spite of their large distances, can jump the 
apparent-magnitude limit of the catalogue. While one looks further and 
further out one finds more and more luminous objects. The consequence 
for an apparent--magnitude catalog (which is the case with galaxy 
catalogues) is that dwarf galaxies fade out quickly with distance, and 
finally at the largest distances the extremely luminous and rare 
galaxies are the only ones which can enter the catalogue. 
  
In addition to effects (B1) and (B2), some noise 
 spikes may appear due to gravitational clustering of objects. The large scale 
 distribution of galaxies shows a variety of landscapes containing 
 voids, walls, filaments and clusters in a complex 3D sponge-like 
 pattern.  For instance, since there are many galaxies in clusters, the 
 distances associated to cluster-cluster separations may appear as fake 
 spikes in the {\sl PSH} for galaxies. Typical cluster -- cluster  
 separations are $130 h^{-1}$ Mpc (Guzzo {\em et al.} 1999).  
 Such effects currently occur in 
 N--body simulations that show clustering (Park and Gott 1991). The trouble 
 can be relieved if galaxy clusters instead of galaxies are used as 
 typical objects for probing the topology, although in that case 
 supercluster-supercluster separations might also introduce noise spikes 
 (at a lower level).\\

\section{Numerical implementation} \label{3}

In the following, we perform numerical simulations to evaluate separately the 
magnitudes of the effects listed above. As usual, we 
start from a random distribution of cosmic objects in the fundamental 
domain. In a first step we generate a complete catalogue of 
comoving objects by unfolding the distribution in the universal covering space.  
We refer to this catalogue as the {\it ideal catalogue} 
since it would correspond to ideal observations. In a second step we  
introduce the various errors $(A_{i})-(B_{i})$ in order to build more realistic 
 catalogues which depart from the ideal one.

\subsection{Uncertainties on positions}

To study numerically the errors on the positions, we first assume that  
the cosmological parameters are known with good enough accuracy, so that we  
do not discuss (A3). Then we perform the following calculations.  
\begin{enumerate}  
\item To evaluate the effect of the observational imprecision, we give to each object of the  
ideal catalogue (built on the assumption that the objects are strictly comoving)  
a redshift error to be added to its ideal redshift. We  
assume that the distribution of the redshift error is Gaussian, with mean  
value $\bar z = 0$ and  dispersion $\Delta z$.   
   
Note that, since $\Delta z$ is absolute, the relative error will be less important  
 when we deal with catalogues of higher redshift objects.  
  
\item To discuss (A2), we associate a peculiar velocity to each point  
of the catalogue before unfolding.  Hence, there will be correlations  
between the velocities of two topological images.  It follows that,  
assuming that the time evolution of the peculiar velocity is small,  
the observed velocities of two topological images of the same object  
can have different directions but similar magnitudes (see e.g.   
Roukema and Bajtlik 1999).  As in the previous case, we assume that  
the velocity distribution is Gaussian.  Moreover, velocities at  
different points of space are assumed to be uncorrelated both in  
magnitude and direction, which is indeed not strictly the case in real  
data since large scale streaming motions of galaxies have been  
observed (Strauss and Willick 1995). 
\end{enumerate}

More precisely, to generate the catalog of topological images taking into account the peculiar velocities, we proceed as follows: 
\begin{enumerate} 
    \item We generate a random collection of points $M_{i}$ (named 
 original sources) uniformly distributed inside the fundamental 
 domain. Each point is assigned a velocity $v_{i}$ according to a Gaussian 
 distribution with mean $\bar v$ and dispersion $\Delta v$. 
    \item We unfold this catalogue of original sources to  
    obtain the set of points $M_{k,i}$, images of $M_{i}$ by the  
    holonomies $g_{k} \in \Gamma$. 
    \item Each image $M_{k,i}$ has a redshift $z_{k,i}$ corresponding to a  
    look--back time $\tau_{k,i} \equiv \tau(z_{k,i})$ calculated with formula  
    (\ref{tau_de_z}). 
    \item The final catalogue accounting for the peculiar velocities of 
    the original sources is obtained by applying the $g_{k}$ to $M_{i} + 
    v_{i}\times \tau_{k,i}$ for all couples $(k,i)$. 
\end{enumerate} 
 
Since $\Delta z$ can be interpreted either as an uncertainty on the redshift determination   
or as due to a radial peculiar velocity, both errors (A1) and (A2) can  
be investigated with the same calculations. The interpretation of $\Delta  
z$ in terms of a radial peculiar velocity is useful to compare the orders of magnitude of  
(A1) and (A2). In real data, the error due to (A2) is expected to be greater  
than the error due to (A1). 
 
\subsection{Catalogue incompleteness}  
  
Starting from an ideal catalogue, we simulate two kinds of  
incomplete catalogues:  
\begin{enumerate}  
\item To account for effect (B1), we randomly throw out $p\%$ of the  
objects from the ideal catalogue. In our various runs we vary $p$ while keeping  
approximately constant the  
number of catalogue objects, which means that we must  
increase the number of ``original" objects in the fundamental domain when  
$p$ is increasing.   
\item To simulate effect (B2), we generate a catalogue limited in solid  
angle by selecting only the objects that lie in a beam of aperture  
$\theta$.  The sky coverage is related to $\theta$ via  
\begin{equation}  
q\% = \frac{1}{2}\left(1-\cos(\theta/2)\right).  
\end{equation} 
Again, we keep constant the number of objects in the catalogue when we  
vary $\theta$. Table  
\ref{coverage} below gives typical numbers. 
\end{enumerate}

\begin{table} 
\caption{The multiplication factor, i.e. the ratio  between the total number of entries  
in the catalogue and the number of  
 ``original" objects randomly distributed in the fundamental  
domain, is given as a function of the solid angle of the survey. We have  
also indicated the corresponding sky coverage.}  
\label{coverage} 
\begin{center}  
\begin{tabular}{|c|c|c|}  
\hline  
$\theta $& multiplication factor & sky  
coverage ($q$\%)\\  
\hline  
160$^{\rm o}$ & 6.75 &  41.3 \\  
80$^{\rm o}$ & 1.92 &  11.7 \\  
70$^{\rm o}$ & 1.45 &  9.0 \\  
60$^{\rm o}$ & 1.06 & 6.7\\ \hline  
\end{tabular}  
\end{center} 
\end{table} 
 
As a matter of fact, more than the aperture angle, the depth of the  
survey will be critical for the multiplication factor, since within a beam of given angle, if the  
redshift cut-off is great enough to encompass a distance (in the  
universal covering space) N times greater than the size of the  
fundamental domain, at least N topological images will be expected in  
the beam.

\section{General Results}  \label{4} 
  
\subsection{Euclidean spaces}

We first consider the Euclidean case and we apply the {\sl PSH} 
crystallographic method as described in Lehoucq {\em et al.} (1996), 
restricting to a typical situation where $\Omega_0=0.3$ and 
$\Omega_{\Lambda0} = 0.7$. We choose the topology of the universe to 
be a cubic 3--torus $T_1$ (see e.g. Lachi\`eze--Rey and Luminet (1995) 
for description), with identification length $L = 3,000\,{\rm Mpc}$ for 
a Hubble parameter $H_{0} = 75\,{\rm km.s}^{-1}{\rm Mpc}^{-1}$. In the 
various runs, the number of objects in the catalogue is kept constant 
at $8500$ (this is the order of magnitude of the number of objects in 
current quasar catalogues). We examine separately the effects of 
errors in position due to redshift uncertainty $\Delta z$ and peculiar 
velocities $\Delta v$, and the effects of catalogue incompleteness due 
to selection effects and partial sky coverage. Each of these effects 
will contribute to spoil the sharpness of the topological signal.  For 
a given depth of the catalogue, namely a redshift cut-off $z_{cut}$, 
we perform the runs to look for the critical value of the error at 
which the topological signal fades out. 
 
\begin{figure}   
\centerline{  
\epsfig{figure=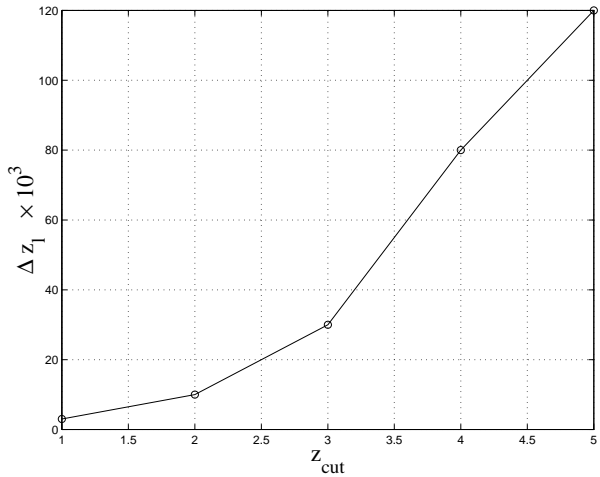,width=8cm}}   
\caption{Plot of $\Delta z_{l}$ as a function of the depth of the
catalogue, for a cubic hypertorus of size $3,000$~Mpc which corresponds
to 45\% of the Hubble radius ($4,950\,h^{-1}$~Mpc) when $\Omega_0=0.3$,
$\Omega_{\Lambda0}=0.7$ and $h=0.75$,
using the {\sl PSH} method.}
\label{plot_deltaz}   
\end{figure}

Figure \ref{plot_deltaz} gives the critical redshift error  $\Delta  
z_{l}$ above which the topological spikes disappear.  
 
The effect of peculiar velocities is very weak, since we find that the 
peculiar velocity must exceed $\Delta v_{l} = 10,000\,{\rm km.s}^{-1}$ 
when $z_{cut} = 1$ and $\Delta v_{l} =40,000\,{\rm km.s}^{-1}$ when 
$z_{cut} = 5$ in order to make the topological signal disappear. As 
already pointed out in Lachi\`eze--Rey and Luminet (1995), the {\it 
observed} peculiar velocities of galaxies have typical values much 
less that $\Delta v_{l}$.

The effects of catalogue incompleteness are summarized in tables  
\ref{t2} and \ref{t3}. The topological signal would be destroyed only  
for a very large rejection percentage or a small aperture angle.  
  
\begin{table} 
\caption{Values of the critical percentage of rejection above which the  
{\sl PSH} spikes disappear, as a function of the redshift cut--off.}  
\label{t2} 
\begin{center} 
\begin{tabular}{|c||c|c|c|} 
\hline 
$z_{\rm cut}$ & $\qquad1\qquad$ & $\qquad2\qquad$ & 
$\qquad\ge 3\qquad$  \\ 
\hline 
$p_{l} [\%]$ & $\qquad70\qquad$ & $\qquad80\qquad$ & 
$\qquad>90\qquad$\\ 
\hline 
\end{tabular} 
\end{center} 
\end{table} 
 
\begin{table} 
\caption{Values of the aperture angle below  
which the {\sl PSH} spikes disappear, as a function of the redshift cut--off.}  
\label{t3} 
\begin{center} 
\begin{tabular}{|c||c|c|c|c|} 
\hline 
$z_{\rm cut}$ & $\qquad2\qquad$ & $\qquad3\qquad$ & 
$\qquad4\qquad$  & $\qquad5\qquad$  \\ 
\hline 
$\theta_{l}$ & $\qquad110^{\rm o}\qquad$ & $\qquad80^{\rm o}\qquad$ & 
$\qquad70^{\rm o}\qquad$ & $\qquad60^{\rm o}\qquad$ \\ 
\hline 
\end{tabular} 
\end{center} 
\end{table}

\subsection{Hyperbolic spaces} 
We now turn to universes with hyperbolic spatial sections and apply 
the {\sl CCP} method as described in Uzan {\em et al.} (1999a). The 
obtention of a topological signal (the so--called {\sl CCP} index) 
strongly depends on the correct determination of the cosmological 
parameters. In Uzan {\em et al.} (1999a) we discussed the problems 
arising from the spanning of the cosmological parameters space with a 
required ``accuracy bin" $\varepsilon$. Latest independent constraints 
on these cosmological parameters from the cosmic microwave background 
(de Bernardis {\em et al.} 2000), the study of supernovae (Efstathiou 
{\em et al.} 1999), of large scale structure at $z = 2$ (Roukema and  
Mamon, 2000)  and of gravitational lensing (Mellier 1999), make us 
hope that we can restrict further the parameters space to apply 
efficiently the {\sl CCP} method. In the following, we assume that the 
cosmological parameters are known with the required accuracy. 
  
We choose the topology of the universe to be described by a Weeks  
manifold (Weeks, 1985) (see also Lehoucq {\em et al.} (1999) for the numerical  
implementation of this topology), assuming the cosmological  
parameters given by $\Omega_0=0.3$ and $\Omega_{\Lambda0}=0$. In  
such a case the number of copies of the fundamental domain within the  
horizon is about 190. However our simulated catalogues are much  
smaller than the horizon volume. We fix the number of objects to  
$1300$, which is a compromise between a realistic  catalogue  
population and a reasonable computing time. 
  
Again we perform the tests by varying the errors $\Delta z$ and  
$\Delta v$ around their average values $\bar z = 0$ and $\bar v = 0$  
until when the {\sl CCP} index falls down to noise level.

Contrarily to the Euclidean case, both slight changes in redshift and in 
 peculiar velocity induce errors in position which dramatically 
 eradicate the topological signal: an error of only $\sim 50\,{\rm 
 kms}^{-1}$ in velocity and an error in redshift of the order of the bin 
 accuracy $\varepsilon \sim 10^{-6}$ drown the {\sl CCP} index into 
 noise. 
  
\begin{figure}   
\centerline{  
\epsfig{figure=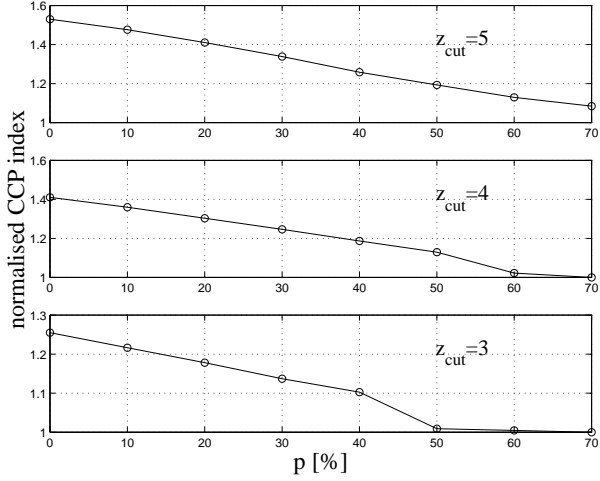,width=8cm}}   
\caption{Plot of the {\sl CCP}--index (normalized to background noise) as a function  
of the rejection percentage of objects $p_{l}$, for various values of  
of the depth of the catalogue.  The topological signal 
disappears when the rejection percentage is greater than $p_{l}$.} \label{plot_ppc}   
\end{figure}

\begin{figure}   
\centerline{  
\epsfig{figure=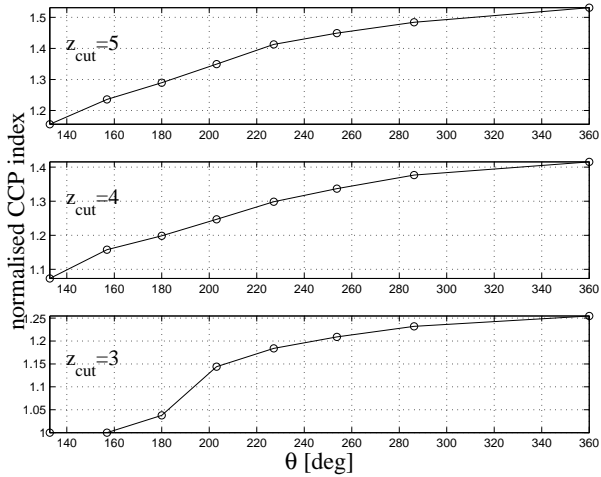,width=8cm}}   
\caption{Plot of the {\sl CCP}--index as a function  
of $\theta_{l}$ for various values of the depth of the catalogue.  
The topological signal disappears when the aperture angle falls down  
below $\theta_{l}$.}   
\label{plot_skycov}   
\end{figure}

The incompleteness effects are less dramatic, as shown in figures 
\ref{plot_ppc} and \ref{plot_skycov}. Again, for a given redshift 
cut--off, we performed the runs by varying $p$ and $\theta$ in order 
to find the critical values at which the topological signal disappears.

\section{Conclusions and perspectives}  \label{5}

Our numerical results have now to be compared with the precisions of 
present experimental 3D data, and to the performances of observational 
programs started or expected to be achieved in the next decade. 
 
At present day, a typical precision   
practical for the spectroscopic uncertainty   
is $\Delta z \sim 0.001$ for an object such as a quasar.  
In clusters, spectroscopic redshifts can be found  
very precisely for individual galaxies. 
 
Concerning peculiar velocities, the typical dispersion velocity is  
$1000 \, \mathrm{km.s}^{-1}$ in rich clusters.  The X-ray velocity of  
the peak of the X-ray distribution would also provide a way to  
estimate the true cluster redshift, including the peculiar velocity of  
the cluster as a whole.  For a quasar, a conservative upper limit to  
the peculiar velocity, assuming the quasar to be at the centre of a  
galaxy, can be taken as $\Delta z \sim 0.002$.  So, from an  
experimental point of view, the uncertainties on the redshifts will be  
dominated by effect (A2), namely the peculiar velocities, rather than  
by the spectroscopic imprecision. 
 
The main limitation of present 3D samples is the small volume of 
existing redshift data.  Future surveys will significantly improve 
both the redshift cut--off and the sky coverage. For instance, the 
Sloan Digital Sky Survey (SDSS) (Loveday 1998) will map in detail 
one-quarter of the entire sky, determining the positions and absolute 
brightnesses of more than 100 million celestial objects. It will also 
measure distances to more than a million galaxies and quasars. More 
precisely, SDSS will map a contiguous $\pi$ steradians area in the 
north Galactic cap, up to a limiting magnitude 23 for two thirds of 
the observing time, together with three southern stripes centred at RA 
$\alpha = 5^{\circ}$ and with central declinations of $\delta = + 
15^{\circ}, 0^{\circ}$ and $-10^{\circ}$ for the remaining one third 
of the time.  Concerning the distance determinations, $10^{6}$ 
galaxies and $10^{5}$ quasars will be observed spectroscopically with 
a resolution $\Delta z/z \sim 5.  10^{-4}$. The main galaxy sample 
will consist of $\sim 900,000$ galaxies up to magnitude 18, with a 
median redshift $z \sim 0.1$. A second galaxy sample will consist in 
$\sim 100,000$ luminous red galaxies to magnitude 19.5 with a median 
redshift $z \sim 0.5$.  Precision on redshifts can be estimated for 
the reddest galaxies to $\Delta z \sim 0.02$.  A sample of $\sim 
100,000$ quasars will be observed, an order of magnitude larger than 
any existing quasar catalogue.  The complete survey data will become 
public by 2005. We can also mention the ESO--VLT Virmos Deep Survey 
Project (Lef\`evre 2000), a comprehensive imaging and redshift survey 
of the deep universe based on more than 150 000 redshifts. 
 
In the field of X--ray observations, the XMM satellite (Arnaud 1996) 
will provide deep insight on X--ray galaxy clusters and active 
galactic nuclei. Also, the XEUS project under study by ESA (Parmar 
{\em et al.} 1999) will be a long--term X--ray observatory at 1 keV, 
with a limiting sensitivity around 250 times better than XMM, allowing 
XEUS to study the properties of galaxy groups at $z = 2$ and active 
galactic nuclei at $z \lesssim 3$.

In the present paper we have investigated how the various observational uncertainties  
will spoil the topological signal expected to arise in the ideal  
situation when crystallographic methods are applied to complete catalogues of  
perfectly comoving objets with zero proper velocities and whose 3D--positions  
are known with infinite accuracy. By numerical simulations we have introduced  
random errors for each possible uncertainty, and we varied the parameters to  determine the limits at which the topological signal vanishes. 
 
Our numerical calculations of the spoiling effects due to the various  
uncertainties $(A_{i})-(B_{i})$ clearly show that the crystallographic  
methods are stable (in the sense that the topological signal is robust  
when data depart from the ideal ones) in the Euclidean case, but highly unstable in the  
hyperbolic case. Indeed in a small multi--connected flat space, realistic  
values of  
peculiar velocities of objects, errors in redshift determinations and  
partial sky coverage will not make the {\sl PSH}  
method to fail. This can be understood by the fact that the  
topological images of a given object are related together by  
Clifford translations which enhance the topological signal.  
On the contrary, in a compact hyperbolic model, holonomies are not Clifford  
translations. The topological signal, built as a {\sl CCP} index, is destroyed  
as soon as small errors are introduced in the position of objects, due  
either to peculiar velocities or to redshift measurement imprecision. 
 
The same kind of critical analysis should be made with the 2D 
topology--detecting methods based on the analysis of CMB data. For 
instance, in the pairs of matched circles method (Cornish {\em et al.} 
1998), it would be necessary to investigate how deviations to the 
ideal situation, such as a non zero thickness of the last scattering 
surface or peculiar motions of the emitting primordial plasma regions, 
would alter the pattern of perfectly matched circles. 
 
To conclude, let us comment on the future of observational cosmic 
topology, at the light of observational constraints on the curvature 
parameters recently provided by BOOMERanG and MAXIMA balloon measurements  
(de Bernardis 
{\em et al.} 2000, Hanany {\em et al.} 2000). Under specific assumptions such as a cold dark 
matter model and a primordial density fluctuation power spectrum, the 
range of values allowed for the energy-density parameter $\Omega = 
\Omega_{0} + \Omega_{\Lambda0}$ is restricted to $ 0.88 < \Omega < 
1.12$ with $95 \%$ confidence. This means that the curvature radius of 
space is as least as great as the radius of the observable 
universe (delimited by the last scattering surface). Such results, if 
accepted, leave open all three cases of space curvature as well as 
most of multi--connected topologies. 
 
A strictly flat space is quite improbable.  Even inflationary scenarios 
predict a value of $\Omega$ asymptocally close to 1, but not strictly 
equal. From a topological point of view, a strictly Euclidean space 
would be interesting since we have shown that the {\sl PSH} method is 
robust enough to provide a topological signal even when realistic 
uncertainties on the data are taken into account. 
 
For compact hyperbolic spaces, if we accept the recent observational  
constraints $ 0.88 < \Omega_{0} + \Omega_{\Lambda0} < 1 $, a number of  
spaceforms such as the Weeks or the Thurston 
manifolds still have topological lengths smaller  
than the horizon size (Weeks: SnapPea). However the topological lens  
effects would be weaker in 
spaceforms with $\Omega$ close to 1 than in spaceforms with $\Omega 
\sim 0.3$ (see figs. 2 and 3 of Lehoucq {\em et al.} 
1999). Furthermore, only the {\sl CCP} method can be applied for 
detecting the topology, and the present article shows that the 
topological signal will fall to noise level as soon as uncertainties 
smaller than the experimental ones are taken into account in the 
simulations. 
 
Eventually, elliptical spaceforms appear to be the most interesting 
case.  On a theoretical point of view, as far as we know, no inflationary model is able to 
drive the density parameter to a value greater than 1, so that if space 
happened to be really elliptical, new models should be built in order to 
explain, e.g., the primordial fluctuations spectrum or the horizon 
problem.  From a topological point of view, since the volumes of (all closed) 
spaceforms are not bounded below (the order of the holonomy group  
can be arbitrarily large), one can always find an elliptical  
space which fits into the Hubble radius even if $ 1 < \Omega_{0} + \Omega_{\Lambda0} < 
1.12 $. On the other hand, the holonomies of such spaces are Clifford 
translations, so that we can hope to apply the robust {\sl PSH} method 
to detect a topological signal. This will be the purpose of our 
subsequent paper.

\vskip0.5cm  
\noindent{\bf Acknowledgements:} It is a pleasure to thank N. Aghanim,   
R. Juszkiewicz, Y. Mellier and B.  
Roukema                               for discussions.  
 
\bibliographystyle{astron} 
 
\end{document}